\documentclass[conference]{IEEEtran}
\usepackage{makecell}
\usepackage{cite}
\usepackage{amsmath,amssymb,amsfonts}
\usepackage{algorithmic}
\usepackage{graphicx}
\usepackage{textcomp}
\usepackage{xcolor}
\usepackage{url}
\usepackage{subfigure}
\usepackage{colortbl}
\usepackage{tabulary}
\usepackage[numbers]{natbib}
\usepackage{etoolbox}
\usepackage{multirow}
\usepackage{hyperref}

\def\BibTeX{{\rm B\kern-.05em{\sc i\kern-.025em b}\kern-.08em
    T\kern-.1667em\lower.7ex\hbox{E}\kern-.125emX}}

\begin{document}

\title{S2AND: A Benchmark and Evaluation System for Author Name Disambiguation}

\author{\IEEEauthorblockN{Shivashankar Subramanian\IEEEauthorrefmark{2}\IEEEauthorrefmark{1},
Daniel King\IEEEauthorrefmark{2}, Doug Downey\IEEEauthorrefmark{2}\IEEEauthorrefmark{3} and
Sergey Feldman\IEEEauthorrefmark{2}}
\IEEEauthorblockA{\IEEEauthorrefmark{2}Allen Institute for AI,
Seattle, WA, USA\\
\IEEEauthorrefmark{3}Northwestern University, Evanston, IL, USA 
\IEEEauthorrefmark{1}University of Melbourne, Australia \\
Email: shivashankar@student.unimelb.edu.au,
daniel@allenai.org,
dougd@allenai.org,
sergey@allenai.org}}

\maketitle

\begin{abstract}
Author Name Disambiguation (AND) is the task of resolving which author mentions in a bibliographic database refer to the same real-world person, and is a critical ingredient of digital library applications such as search and citation analysis.  While many AND algorithms have been proposed, comparing them is difficult because they often employ distinct features and are evaluated on different datasets.

In response to this challenge, we present S2AND, a unified benchmark dataset for AND on scholarly papers, as well as an open-source reference model implementation.  Our dataset harmonizes eight disparate AND datasets into a uniform format, with a single rich feature set drawn from the Semantic Scholar (\texttt{S2}) database.  Our evaluation suite for S2AND reports performance split by facets like publication year and number of papers, allowing researchers to track both global performance and measures of fairness across facet values.

Our experiments show that because previous datasets tend to cover idiosyncratic and biased slices of the literature, algorithms trained to perform well on one on them may generalize poorly to others.  By contrast, we show how training on a union of datasets in S2AND results in more robust models that perform well even on datasets unseen in training.  The resulting AND model also substantially improves over the production algorithm in \texttt{S2}, reducing error by over 50\% in terms of $B^3$ F1.
We release our unified dataset, model code, trained models, and evaluation suite to the research community.\footnote{\url{https://github.com/allenai/S2AND/}}

\end{abstract}


\section{Introduction}
A central challenge in curating large bibliographic databases is determining which author mentions in the database refer to the same real-world person.  The problem is challenging because distinct authors often share the same name, and likewise the same author may appear under multiple distinct names.
Major bibliographic databases struggle to attribute papers to the correct authors, and authors often resort to manually correcting the databases as a labor-intensive work-around.
Automatically determining which {\em author records}, i.e. author name strings assigned to a given paper, refer to the same real-world person is a task known as Author Name Disambiguation (AND).  Accurate AND is critical for features such as searching or browsing publications by author, reporting up-to-date author profiles with accurate bibliometrics, and other capabilities.

A variety of AND algorithms have been introduced and evaluated in previous work \cite{ferreira2012brief}.  In general, these algorithms operate in three steps: first, they heuristically group candidate duplicate records into blocks for tractability, based on names; then, they score the similarity of each pair of records within a block based on features such as affiliations, co-authors, and paper content; and finally, they cluster the records based on the pairwise scores.  Despite considerable progress, it is difficult to accurately assess the effectiveness of today's AND techniques from the existing literature.  Comparing the algorithms to one another is difficult because they often have different sets of features, and are evaluated on different datasets (see Table \ref{tab:data-prop}). Further, existing AND datasets cover biased portions of the literature \cite{vishnyakova2019new}, e.g. the popular Aminer dataset \cite{zhang2018name} consists of only Chinese names, whereas SCAD-zbMATH \cite{muller2017data} contains only mathematics papers. Thus, it is unclear whether a method that performs best on one AND dataset will provide strong performance on other datasets or the diverse data found in bibliographic databases in practice.

In this paper, we present S2AND, a new author disambiguation dataset that combines eight previous datasets into a single resource with a uniform format and a consistent, rich feature set.  S2AND's feature values are obtained by aligning its author records to the Semantic Scholar (S2) bibliographic knowledge graph.  In experiments with a new AND model that is representative of the state-of-the-art, we show how training the model on the union of datasets in S2AND improves AND accuracy.  Training on S2AND provides advantages in both the {\em in-domain} setting, in which the training set for the target test set is included in training, and the {\em out-of-domain} setting, where it is not.  In the out-of-domain setting, training on the union of datasets in S2AND achieves or equals the top clustering accuracy in 4/7 experiments, compared to at most 1/7 achieved when training on any of the original datasets alone.
Further, we find that our S2AND-trained system provides dramatic improvements over the author disambiguation algorithm in production on \texttt{S2} today, reducing error by more than 50\% in terms of $B^3$ accuracy.  

Thus, in addition to serving as a benchmark for comparing AND algorithms, S2AND provides training data that may improve the accuracy and generalization of AND systems in practice.

To summarize, our contributions are:
\begin{enumerate}
    \item S2AND, a new training dataset and evaluation benchmark for author name disambiguation that unifies and extends previous resources,
    \item A new open-sourced reference AND algorithm implementation which is comparable in performance to state-of-the-art, 
    \item Experiments showing that training on S2AND improves generalization compared to the single-dataset approach taken in previous work, and
    \item A comparison against the existing Semantic Scholar production system, showing that the S2AND-trained system performs better across all the datasets, including across all ranges for all facets examined, such as the number of papers by the author or the publication date.
\end{enumerate}

\begin{table*} [ht]
\centering
\caption{\label{tab:data-prop} Features provided by previous AND datasets (Ref=References, Abs=Abstract, Afl=Affiliation). The PubMed and Medline datasets provide only records (author name and PMID of paper) and their grouping. Datasets after alignment to \texttt{S2} have all features available in S2.}
 \begin{tabular}{l c c c c c c c c c} 
 \hline
 Dataset & Co-author &	Ref	& Title	 & Abs	& Position & Year &	Afl & Email	& Venue \\ [0.5ex] 
 \hline
 Aminer \cite{zhang2018name} & \checkmark &  & \checkmark & \checkmark & \checkmark & \checkmark & \checkmark &  & \checkmark \\ 
 Arnetminer \cite{tang2011unified} & \checkmark & \checkmark & \checkmark &  & \checkmark & \checkmark & \checkmark &  & \checkmark \\ 
 INSPIRE \cite{louppe2016ethnicity} & \checkmark & \checkmark & \checkmark & \checkmark & \checkmark & \checkmark & \checkmark &  & \checkmark\\
 KISTI (DBLP) \cite{kang2011construction} & \checkmark &  & \checkmark &  & \checkmark & \checkmark &  &  & \checkmark \\ 
 Medline \cite{vishnyakova2019new} &  &  &  &  &  &  &  &  &  \\
 PubMed \cite{song2015exploring} &  &  &  &  &  &  &  &  &\\
 QIAN (DBLP) \cite{qian2015dynamic} & \checkmark &  & \checkmark & \checkmark &  & \checkmark &  &  & \checkmark \\ 
 SCAD-zbMATH \cite{muller2017data} & \checkmark &  & \checkmark &  & \checkmark & \checkmark &  &  & \checkmark \\ [1ex] 
 \hline
\end{tabular}
\vspace{-15pt}
\end{table*}

\section{Previous work}
\label{sec:existing}

Author Name Disambiguation has been studied for some years, which has led to the introduction of a variety of datasets differing along important dimensions, such as scientific domain, author ethnicity, feature availability, method of curation, and degree of ambiguity.  Early datasets included Arnetminer \cite{wang2011adana}, a small manually-curated set of highly ambiguous author names from papers in a variety of domains, and KISTI \cite{kang2011construction}, a larger set of much less ambiguous names from computer science papers mined from DBLP \cite{qian2015dynamic}. Later datasets focused on different individual domains, such as PubMed in the medical domain \cite{song2015exploring}, INSPIRE in high-energy physics \cite{louppe2016ethnicity} and SCAD-zbMATH in mathematics \cite{muller2017data}.  Different datasets also differ with respect to geographic name origin---for example, Aminer \cite{zhang2018name} contains predominantly Chinese names, whereas SCAD-zbMATH and INSPIRE \cite{louppe2016ethnicity} contain mostly Western names. 
Different datasets also include different features (author name, affiliation, email, paper title, journal details, keywords, coauthor lists, domain-specific features like MeSH indicators, etc.) with varying levels of availability (see Table \ref{tab:data-stats}). 
\citet{vishnyakova2019new} noted the bias of existing AND datasets, and argued that the datasets were not representative of the challenge faced by scholarly databases, which often lack the datasets' most valuable features (e.g., author affiliation in Pubmed).  That work developed the Medline dataset from the medical domain, and showed that models trained on PubMed failed to generalize to Medline, highlighting the practical significance of the dataset bias.  Our goal is to unify this previous body of work and present a single AND resource that, because it is comprised of the union of disparate datasets, covers substantially more of the true diversity of the AND task faced in practice. 

Several classification models have been
used for learning the pairwise similarity function, including Naive Bayes \cite{han2004two}, Logistic Regression \cite{levin2012citation}, Support Vector Machines \cite{huang2006efficient,han2004two,song2015exploring}, Decision Trees (C4.5) \cite{vishnyakova2019new}, Random Forests (RF) \cite{treeratpituk2009disambiguating,louppe2016ethnicity,song2015exploring,10.1145/3383583.3398568,levin2012citation}, Deep Neural Networks (DNN) \cite{tran2014author} and Gradient Boosted Trees (GBT) \cite{kim2018web,louppe2016ethnicity,kim2019hybrid,10.1145/3383583.3398568}.  \citet{tran2014author} use DNNs with manually-crafted features, whereas \citet{atarashi2017deep} leveraged a DNN to learn feature representations from bag-of-words vectors. \citet{zhang2018name} used a DNN to first learn a representation for each record, and refine it using a graph autoencoder, where the graph is constructed based on the similarity between records. Similarly, \citet{kim2019hybrid} used a DNN to learn a representation for each record to provide a similarity feature, and showed that incorporating the feature among others within GBTs outperformed other ways of using neural representations from earlier work \cite{atarashi2017deep,zhang2018name}.
We follow a similar approach in our experiments, using GBT as our classifier over a union of features from previous datasets plus paper representations output by a state-of-the-art deep neural network \cite{cohan-etal-2020-specter}.  Our results show that this approach is representative of the state-of-the-art in AND.

Similarly, a variety of clustering techniques have been used for grouping records based on pairwise distances, including spectral clustering \cite{han5name}, affinity propagation \cite{fan2011graph}, density-based spatial clustering of applications with noise (DBSCAN) \cite{huang2006efficient}, and hierarchical agglomerative clustering (HAC) \cite{liu2014author,louppe2016ethnicity,zhang2018name}. We evaluate the popular choices \cite{ferreira2012brief} of DBSCAN and HAC, and find that HAC performs best.  Exploration of additional clustering methods is an item of future work.

The increasing use of AI systems in society has led to an increased emphasis on the fairness of the systems, in addition to their global predictive accuracy \cite{blodgett2020language}.  If AND systems have different error rates for different groups, this can be harmful because the publication records in scholarly databases are used as an input to decisions regarding hiring, promotion, conference responsibilities, and more. Because AND mistakes can cause representational and allocational harm, and no system will ever be perfect, it is critical that any live AND service easily allows authors to correct mistakes made by the system. Other than using self-reported demographic attributes \cite{bogen2020awareness}, research has focused on using inferred gender from names for studying gender disparities in authorship and citation trends \cite{mohammad2020gender,vogel2012he, MihaljeviBrandt2016TheEO}. Similarly, \citet{bertrand2004emily} used inferred gender and race for studying disparities in hiring. In our work, we use the inferred ethnicity of authors, predicted using Ethnea \cite{torvik2016ethnea}, in addition to their prolificity (based on count of papers published), to evaluate the disparity of AND systems' performance across groups.  To our knowledge, ours is the first study to begin to analyze the fairness of modern AND systems.

In our analysis of performance by ethnicity, we will refer to {\em estimated name geographic origin}. This value is predicted using Ethnea, as mentioned above, and is a highly imperfect measure of actual author attributes. {\em Estimated name geographic origin} is predicted by looking up the likely country for a particular name, and then probabilistically mapping countries to ethnicities. The resulting classes from the Ethnea tool are clearly not comprehensive or representative, as `African' is a single category. Despite these significant shortcomings, we would like to know if our system has performance disparities across the predicted groups of the Ethnea tool. Having no disparity across these groups would {\em not} indicate a perfectly fair model, but having significant disparity across these groups would {\em strongly suggest} that real performance disparities exist, which deserve further investigation. Additionally, while our analysis is performed by mapping individual names to {\em estimated name geographic origin}, we do not release the individual values, only the aggregated performance statistics, due to the possible harm of misidentifying attributes of individuals \citep{Hamidi2018GenderRO, Mihaljevic2019ReflectionsOG}. The gold standard would be to ask authors to self-identify \citep{Hamidi2018GenderRO}, but this is unfortunately not possible at scale, and suffers from other issues, including reporting bias \citep{Mihaljevic2019ReflectionsOG}. We hope that future work continues to address representational harm in bibliographic databases and its impact on people and their careers.\footnote{We did not observe any difference across stereotypically estimated name gender (also from the Ethnea tool), but refrain from including detailed results because the estimated gender is not of sufficient quality, not self-reported, and binary.}

\vspace{-10pt}
\section{S2AND}
\label{sec:s2and-data}

We now detail how we constructed the S2AND benchmark dataset by aligning existing resources with Semantic Scholar.

\begin{table*} [ht]
\begin{small}
\caption{\label{tab:data-stats} Statistics on the S2AND dataset.  For each source dataset, the first four columns list the number of data objects from that dataset in S2AND (with the corresponding count from the original dataset in parentheses).  The remaining columns list the prevalence of each feature in S2AND.  First name, affiliation and email coverage are expressed as proportions of records, and the other coverages are proportions of papers. First name denotes the proportion of records containing a full first name.}
\setlength\tabcolsep{4pt} % default value: 6pt
\scalebox{0.85}{%
 \begin{tabular}{l r r r r c c c c c c} 
 \hline
 Dataset & \#Blocks & \#Records & \#Papers&	\#Clusters&	Abstract&	References&	First name &	Affiliation& 	Email & Venue \\ [0.5ex] 
 \hline
 Aminer  & 2096 (600) & 157,448 (208,827) & 153,147 (203,078) & 31,848 (39,781) & 0.9599 & 0.4344 & 0.9346 & 0.7375 & 0.0199 & 0.7009 \\ 
 Arnetminer  & 130 (103) & 7,144 (7,528) & 7,067 (7,447) & 1,512 (1,726) & 0.9420 & 0.7164 & 0.9730 & 0.7424 & 0.0000 & 0.9716 \\ 
 INSPIRE & 21,477 (12,458) & 536,564 (1,201,763) & 265,497 (360,066) & 14,996 (15,388) & 0.9137 & 0.3239 & 0.4997 & 0.5576 & 0.0009 & 0.2835\\
 KISTI & 1044 (881) &  40,383 (41,673)  & 36,447 (37,613) & 6,856 (6,921) & 0.9580 & 0.7840 & 0.9731 & 0.7890 & 0.0000 & 0.9972 \\ 
 Medline & - & 3,738 (3,750) & 3,707 (3,744) & - & 0.8778 & 0.4715 & 0.6027 & 0.6169 & 0.0594 & 1.0000 \\
 PubMed & 45 (41) & 2,871 (2,875) & 2,871 (2,875) & 385 (385) & 0.9516 & 0.4932 & 0.7886 & 0.9436 & 0.4720 & 1.0000\\
 QIAN & 606 (580) & 6,542 (6,717) & 6,542 (6,717)  & 1,188 (1,201) & 0.9590 & 0.7733 & 0.9934 & 0.7634 & 0.0000 & 0.9954 \\ 
 SCAD-zbMATH & 2,260 (2,136) & 15,181 (33,810) & 12,289 (28,321) & 2,334 (2,946) & 0.5109 & 0.1501 & 0.6717 & 0.3415  & 0.0000 & 0.3728\\ [1ex] 
 \hline
\end{tabular}}
\end{small}
\vspace{-10pt}
\end{table*}

\begin{table} [ht]
\begin{small}
\caption{\label{tab:features} Features used for constructing a pairwise-linkage model. Name counts were obtained from the entire Semantic Scholar corpus.}
\scalebox{0.80}{%
 \begin{tabular}{l l} 
 \hline
 Feature & Similarity Functions \\ [0.5ex] 
 \hline
 First name & equal or not,  full first name or initials \\
 & prefix distance, levenshtein distance \\
 & longest common subsequence distance \\
 & jaro winkler similarity \\
 Middle name & Jaccard overlap of initials, equal or not \\
 & missing or not, full middle name or initials \\
 Affiliation & 1-3 word n-gram Jaccard overlap \\
 Email & prefix equal or not, suffix equal or not \\
 Co-authors & Jaccard overlap of block keys \\
 & Jaccard overlap of full names \\ 
 & 2-4 char n-gram Jaccard overlap \\
 Venue & 2-4 char n-gram Jaccard overlap \\
 Year & absolute difference \\
 Title & 1-3 word n-gram Jaccard overlap \\
 & 2-4 char n-gram Jaccard overlap \\
 References & 2-4 char n-gram Jaccard overlap of author names \\
 &  2-4 char n-gram Jaccard overlap of titles \\
 &  2-4 char n-gram Jaccard overlap of venues, journals \\
 & 2-4 char n-gram Jaccard overlap of author block keys \\
 & whether either of the two papers cite each other \\
 & Jaccard overlap of the references (co-citation) \\
 Author position & absolute difference \\
 Abstract & count of papers with abstracts (0, 1, or 2) \\
 Language & count of English papers, same language or not \\
 & count of identified languages \\
 Name counts & min. count of first, first+last, last, first initial+last \\
 & max. count of first, first+last \\
 SPECTER embeddings & cosine similarity \\
 Journal & 2-4 char n-gram Jaccard overlap \\ [1ex] 
 \hline
\end{tabular}}
\end{small}
\vspace{-10pt}
\end{table}

S2AND is a union of eight existing AND datasets, detailed below.  We analyzed a few more datasets from the literature, and left out datasets such as Han-DBLP \cite{han5name} due to erroneous cluster assignments \cite{muller2017data}, and REXA \cite{culotta07author}, a small dataset containing 13 blocks, due to typographical errors in paper details which made it prohibitively hard to align with S2. The very recent Author-ity2009 data set \cite{Kim_2021} appeared after our investigation was complete, so it was not considered for inclusion in S2AND. In the following, a {\em record} refers to an author name appearing on a particular paper (aka {\em signature}), which can be characterized by  associated features (like the author's listed affiliation and email) discussed in later sections.  A {\em block} refers to a subset of records identified in the dataset as potentially the same author; different datasets use different blocking functions, but a typical choice is to put all records with the same last name and first initial into the same block.  Except in the case of Medline, all datasets include a ground-truth partition of the records into disjoint clusters that represent the same author. The datasets that comprise S2AND are:

\noindent{\textbf{Aminer}} \cite{zhang2018name}: manually-disambiguated records from Aminer.cn. 

\noindent{\textbf{ArnetMiner}} \cite{tang2011unified}: Ambiguous names gathered from ArnetMiner (a predecessor of Aminer), and manually disambiguated by comparing affiliations, email addresses, and information on personal webpages. 

\noindent{\textbf{INSPIRE}} \cite{louppe2016ethnicity}: Paper records that have been claimed by their original authors on INSPIRE, an online bibliographic database.  Approximately 13\% of the publication records have been claimed by their original authors and are verified automatically based on persistent identifiers or manually by professional curators. 

\noindent{\textbf{KISTI}} \cite{kang2011construction}: Mentions from the DBLP bibliographic database manually disambiguated using Web search queries.  

\noindent{\textbf{Medline}} \cite{vishnyakova2019new}: Randomly sampled Medline publications. This dataset differs from the others in that it is only a pairwise classification dataset, and does not include full ground-truth clusters.

\noindent{\textbf{PubMed}} \cite{song2015exploring}:  First-author records from PubMed papers, annotated by identifying each author's publication list via Web search.  

\noindent{\textbf{QIAN}} \cite{qian2015dynamic}: A union of other AND datasets, including Han-DBLP \cite{han5name} and ArnetMiner \cite{tang2011unified}, manually de-duplicated and corrected for errors.  

\noindent{\textbf{SCAD-zbMATH}} \cite{muller2017data}: Community-curated mathematics papers from the zbMATH bibliographic database.

For unifying the above AND datasets to \texttt{S2}, we use a semi-automated process.  Given a paper, we first perform a text search for its title on Semantic Scholar, and choose the top 10 results.\footnote{In the case of PubMed and Medline, we instead align using the given PubMed IDs and \url{https://api.semanticscholar.org/}.}  We manually verified this approach using QIAN, and found that the top 10 results provided nearly 99\% recall.  We then re-rank the top 10 results based on similarity of other metadata such as number of authors, author names, venue/journal details and year of publication.  In manual inspection, the top-ranked result was a correct match for about 98\% of the query records.  We manually inspect any cases in which the top-ranked result fails to match the author name of the query record, and discard cases in which the names are actually different.  Finally, our different datasets have some overlap, and when two datasets disagree on whether to cluster a pair of records, we exclude the pair from S2AND (manual inspection revealed that no dataset was fully correct).

Not all the papers in the original datasets had a match in the \texttt{S2} database (similar to the outcome in \cite{kim2018evaluating}, where datasets were aligned to DBLP). On average 82\% of the records were aligned, with nearly complete alignment for PubMed and Medline, and less than 50\% for INSPIRE and SCAD-zbMATH which contain Physics and Mathematics publications respectively (see Table \ref{tab:data-stats}). In spite of this, the datasets can support a holistic evaluation of a feature-based AND system, and assess the performance of \texttt{S2}.

The details of the individual datasets after alignment, including the prevalence of different features obtained from \texttt{S2} for each dataset's records, are given in Table \ref{tab:data-stats}.  Due to low coverage of affiliations data in \texttt{S2} at the time of data collection, we supplement with affiliations from the Microsoft Academic Graph (MAG) \cite{conf/www/SinhaSSMEHW15,wang2019a} when \texttt{S2} affiliations are missing.

\section{Reference Author Name Disambiguation System}
\label{sec:s2and-pipeline}

To report baseline results on our benchmark, and evaluate the utility of training on S2AND, we developed a reference AND implementation that is representative of the state-of-the-art.  We release the implementation as part of our benchmark.  It follows the typical three-stage AND approach: (A) blocking, (B) pairwise similarity estimation, and (C) clustering.  We describe each step below.

\subsection{Blocking}
For tractability, records first are partitioned into disjoint, potentially-coreferent blocks.  We put records in the same block {\em iff} they match on the author's first initial and last name.

\subsection{Pairwise Similarity Estimation}
We estimate the similarity of each pair of records in a block using a classifier trained to predict if two records were written by the same author.  We use a gradient-boosted trees (GBT) classifier over the feature set listed in Table \ref{tab:features}, following previous work as discussed in Section \ref{sec:existing}. We restrict to generic features that can be constructed using the information in bibliographic databases, and avoid domain-specific features such as MeSH indicators or those requiring web search. A distinct but rarely used \cite{10.1145/1552303.1552304} feature is name popularity, which we estimate as number of distinct referents of author names across \texttt{S2}.  Another feature is motivated by the success of neural representations to compute similarity features (see Section \ref{sec:existing}).  Specifically, we compute SPECTER embeddings \cite{cohan-etal-2020-specter} for each record using the paper's title and abstract (if available).  SPECTER is a recent document-embedding approach, trained on the citation graph to produce paper embeddings applicable across multiple tasks.

We use LightGBM \cite{ke2017lightgbm} as our pairwise classifier, and the hyperopt package \cite{bergstra2013hyperopt} to tune 11 hyperparameters\footnote{See \href{https://github.com/allenai/S2AND/blob/5ee7b1b282820dbdcb86a46222ea7ea81d9f2fd5/s2and/model.py\#L701}{{\color{blue} \underline{the code}}} for details.} using a held-out set. We impose feature-wise monotonicity constraints on some features to both (a) regularize the model and (b) constrain the model to behave sensibly when faced with data from outside of the training distribution. For example, we constrain the model to (all other features held constant) decrease the output probability of two records being by the same author if the year difference between the two papers increases. 

The final pairwise classifier ensembles two models: the classifier discussed above, and a version of it that we call `nameless', which is identical in every respect but does not have any features related to the surface forms of the author names (co-author names are still included). The reason for this was practical--we observed that the pairwise model over-relied on name features even in the presence of other metadata. While ensembling with the `nameless' model decreased pairwise AUROC performance, it increased the $B^3$ clustering metric and improved qualitative performance (see Table \ref{tab:main_ablation}).

\subsection{Agglomerative Clustering}

Using the trained pairwise classifier from the previous step we construct a distance matrix $D$ where $D_{ij}$ is the probability that two records $i$ and $j$ are not by the same author. We then partition each block into clusters with hierarchical agglomerative clustering \cite{mullner_2013} over the matrix $D$.  The clustering depends on a {\em linkage} function that estimates the dissimilarity between two clusters, in terms of the pairwise distances between the individual elements of each cluster.  Our experiments evaluate several alternatives, and we find that a straightforward average of all the pairwise distances performs best.

We tune only the \verb|eps| hyperparameter for agglomerative clustering, defined as the distance threshold above which clusters will not be merged. This is tuned on a held-out set of blocks using hyperopt.

\section{Experimental Results}

In this section, we first validate that our AND reference implementation is comparable to the state-of-the-art when evaluated on pre-existing AND datasets.  Then, we perform our primary evaluation of training on S2AND, in both the in-domain and out-of-domain setting, showing that training on a union of datasets in S2AND results in more robust performance compared to the standard approach of training on a single dataset.  Finally, we analyze performance in ablation studies and across different facet values, and compare a S2AND-trained system against the AND system used in Semantic Scholar (\texttt{S2}). As there can be variations in names for various reasons, \texttt{S2}'s blocks are not always the same as the blocks provided in the original datasets. To ensure a fair comparison, we use the blocks from the original datasets for state-of-the-art evaluation (Section \ref{sec:sota}), and \texttt{S2}'s blocks for other experiments (Section \ref{sec:s2andexp} and Section \ref{sec:comps2}).

Our evaluation uses two standard metrics.  To evaluate the pairwise classifier's performance in isolation, we use area under the ROC curve (AUROC).  To evaluate the final end-task clustering performance, we use the $B^3$ F1 metric \cite{Bagga98algorithmsfor}, which is the average of F1 scores for each of the individual records. For state-of-the-art comparisons, we use other metrics in order to compare directly with existing literature. These metrics include average precision for pairwise similarity ranking, and pairwise F1 for the final clustering phase \cite{levin2012citation}.

\subsection{Comparison to state-of-the-art performance}
\label{sec:sota}
The goal of our work is not to propose a new state-of-the-art AND algorithm.  However, to ensure that our evaluation of the S2AND dataset is representative of the current state-of-the-art, we start by showing that our reference AND implementation achieves comparable performance to existing state-of-the-art algorithms when evaluated in the same setting considered in previous work, where we train on records from a single dataset and test on held-out data from the same distribution.  We evaluate on the datasets aligned to \texttt{S2} (Table \ref{tab:data-stats}), which retains a majority of the original records, and contains features from \texttt{S2}. We use the evaluation settings from the state-of-the-art approaches, and compare against their reported performance. For Aminer, the dataset provides the training and test block splits used by \citet{zhang2018name} (note that there is a discrepancy in the statistics of the data mentioned in the paper compared to the data publicly released --- 70,258 vs. 203,078 publications, respectively) and \citet{kim2019hybrid}. The Inspire dataset provides records split into training and test sets \cite{louppe2016ethnicity}, and Medline similarly has pre-determined train and test pairwise classification records \cite{vishnyakova2019new}. For PubMed and KISTI, we use the same cross-validation settings mentioned in \citet{vishnyakova2019new} and \citet{santana2015combination}, respectively.\footnote{We do not have access to the precise CV splits used in those works.} For KISTI's state-of-the-art performance, we also include DBLP's performance as evaluated by \citet{kim2018evaluating}.
The results are shown in Table \ref{tab:sota}.  While there is some variance on the Aminer dataset, on average S2AND's performance is comparable to the published state-of-the-art.

\begin{table}
\begin{small}
\caption{\label{tab:sota} Comparison of our reference implementation vs. previously published results.  Our algorithm performs comparably to the previous state-of-the-art.}
 \scalebox{0.95}{%
 \begin{tabular}{l l l l l} 
 \hline
Dataset & Metric & Task & Ours & SoTA\\ [0.5ex] 
 \hline
Aminer & Average Precision & Classification & \textbf{0.758} & 0.691 \cite{kim2019hybrid}\\ [0.5ex] 
 Aminer & Pairwise Macro F1 & Clustering & 0.613 & \textbf{0.678} \cite{zhang2018name}\\ [0.5ex] 
 \hline
 INSPIRE & $B^3$ F1 & Clustering &0.974  &\textbf{0.987} \cite{louppe2016ethnicity}\\ [0.5ex] 
 INSPIRE & Pairwise Macro F1 & Clustering & 0.980 &\textbf{0.989} \cite{louppe2016ethnicity}\\ [0.5ex] 
  \hline
  PubMed & F1 & Classification & \textbf{0.926} & 0.897 \cite{vishnyakova2019new}\\
  \hline
  Medline & F1 & Classification & \textbf{0.901} & 0.872 \cite{vishnyakova2019new}\\
  \hline
  \multirow{2}{*}{KISTI} & \multirow{2}{*}{Pairwise Macro F1} & \multirow{2}{*}{Clustering} & \multirow{2}{*}{\textbf{0.918}} & 0.816 \cite{santana2015combination} \\
  & & & & 0.917 \cite{kim2018evaluating}\\
  \hline
\end{tabular}}
\end{small}
\vspace{-10pt}
\end{table}

\begin{table*}\centering
\makebox[0pt][c]{\parbox{1.025\textwidth}{%
    \begin{minipage}[b]{0.5\hsize}
        \caption{\label{tab:transfer_b3} $B^3$ F1 clustering performance of training on various datasets, evaluated on different target test sets. \texttt{S2} denotes the performance of the production Semantic Scholar system. Italicized gray entries are for the in-domain setting, and un-italicized entries are for the out-of-domain setting. Bold indicates the best out-of-domain result in each column.}
        \centering
        \setlength\tabcolsep{1.0pt}
        \scalebox{0.99}{%
        \begin{tabular}{l c c c c c c c c} 
        \hline
        Train $\downarrow$ / Test $\rightarrow$ & Aminer & Arnetminer & Inspire & Kisti & Pubmed & Qian & Zbmath \\[0.5ex]
        \hline
        Aminer                 & \cellcolor[gray]{0.9}\textit{0.774} & 0.838 & 0.871 & 0.935 & 0.922 & 0.905 & 0.875 \\
        Arnetminer             & 0.688 & \cellcolor[gray]{0.9}\textit{0.872} & 0.946 & 0.939 & 0.892 & 0.936 & 0.926 \\
        Inspire                & 0.557 & 0.771 & \cellcolor[gray]{0.9}\textit{0.961} & \textbf{0.946} & 0.869 & 0.875 & 0.956 \\
        Kisti                  & 0.611 & 0.859 & \textbf{0.952} & \cellcolor[gray]{0.9}\textit{0.954} & 0.902 & 0.928 & 0.955 \\
        Pubmed                 & 0.619 & 0.773 & 0.878 & 0.906 & \cellcolor[gray]{0.9}\textit{0.910} & 0.887 & 0.892 \\%& 0.841 \\
        Qian                   & 0.684 & \textbf{0.871} & 0.945 & 0.934 & 0.920 & \cellcolor[gray]{0.9}\textit{0.944} & 0.942 \\
        Zbmath                 & 0.487 & 0.747 & 0.945 & 0.897 & 0.692 & 0.825 & \cellcolor[gray]{0.9}\textit{0.953} \\
        \texttt{S2}            & 0.533 & 0.729 & 0.743 & 0.934 & 0.816 & \textbf{0.937} & 0.795 \\
        S2AND (-target)        & \textbf{0.710} & 0.866 & 0.936 & 0.943 & \textbf{0.944} & \textbf{0.937} & \textbf{0.963} \\
        S2AND (+target)        & \cellcolor[gray]{0.9}\textit{0.764} & \cellcolor[gray]{0.9}\textit{0.864} & \cellcolor[gray]{0.9}\textit{0.962} & \cellcolor[gray]{0.9}\textit{0.948} & \cellcolor[gray]{0.9}\textit{0.928} & \cellcolor[gray]{0.9}\textit{0.929} & \cellcolor[gray]{0.9}\textit{0.961} \\
         \hline
        \end{tabular}}
    \end{minipage}
    \hfill
    \begin{minipage}[b]{0.5\hsize}
        \caption{\label{tab:transfer_roc} AUROC classification performance for different training sets, evaluated on different target test sets. Italicized gray entries are for the in-domain setting, and un-italicized entries are for the out-of-domain setting. Bold indicates best out-of-domain result in each column.}
        \setlength\tabcolsep{1.0pt}
        \scalebox{0.99}{%
         \begin{tabular}{l c c c c c c c c c} 
         \hline
        Train $\downarrow$ / Test $\rightarrow$ & Aminer & Arnetminer & Inspire & Kisti & Pubmed & Qian & Zbmath & Medline \\[0.5ex] 
         \hline
        Aminer                 & \cellcolor[gray]{0.9}\textit{0.933} & 0.901 & 0.916 & 0.980 & 0.974 & 0.946 & 0.802 & 0.951 \\
        Arnetminer             & 0.883 & \cellcolor[gray]{0.9}\textit{0.917} & 0.889 & 0.975 & 0.966 & 0.950 & 0.799 & 0.953 \\
        Inspire                & 0.927 & 0.903 & \cellcolor[gray]{0.9}\textit{0.959} & \textbf{0.983} & \textbf{0.980} & 0.938 & \textbf{0.868} & 0.952 \\
        Kisti                  & \textbf{0.930} & \textbf{0.934} & 0.910 & \cellcolor[gray]{0.9}\textit{0.984} & 0.979 & 0.956 & 0.824 & 0.948 \\
        Pubmed                 & 0.927 & 0.899 & 0.893 & 0.982 & \cellcolor[gray]{0.9}\textit{0.985} & 0.937 & 0.733 & 0.950 \\               
        Qian                   & 0.899 & 0.910 & 0.864 & 0.975 & 0.964 & \cellcolor[gray]{0.9}\textit{0.952} & 0.789 & 0.953 \\
        Zbmath                 & 0.909 & 0.847 & 0.902 & 0.970 & 0.962 & 0.903 & \cellcolor[gray]{0.9}\textit{0.875} & 0.955 \\
        Medline                & 0.878 & 0.859 & 0.892 & 0.961 & 0.953 & 0.922 & 0.756 & \cellcolor[gray]{0.9}\textit{0.964} \\
        S2AND (-target)        & 0.904 & 0.921 & \textbf{0.929} & 0.981 & 0.970 & \textbf{0.957} & 0.852 & \textbf{0.971} \\
        S2AND (+target)        & \cellcolor[gray]{0.9}\textit{0.917} & \cellcolor[gray]{0.9}\textit{0.929} & \cellcolor[gray]{0.9}\textit{0.954} & \cellcolor[gray]{0.9}\textit{0.981} & \cellcolor[gray]{0.9}\textit{0.980} & \cellcolor[gray]{0.9}\textit{0.958} & \cellcolor[gray]{0.9}\textit{0.882} & \cellcolor[gray]{0.9}\textit{0.972} \\
         \hline
        \end{tabular}}
        
    \end{minipage}
}}
\vspace{-10pt}
\end{table*}

\subsection{S2AND vs. single-data-set training}
\label{sec:s2andexp}
We now turn to our primary evaluation of how training on S2AND impacts AND system performance, compared to the previous approach of training on a single dataset.  We evaluate in both the in-domain and out-of-domain setting. In these experiments, we first split each dataset into train, validation, and test splits based on blocks derived from \texttt{S2} names (80/10/10 split). For training and tuning the pairwise model, we then sample pairs from train and validation respectively, and from test for AUROC evaluation. For fitting and evaluating the clusterer we use the validation and test splits respectively. Each dataset contributes at most 100,000 pairwise examples to training, and at most 10,000 to each of validation and test. We evaluate on disjoint test sets from each of the seven (eight for pairwise evaluation) original datasets.  When using a single dataset for training, testing on test data from the same dataset evaluates in the {\em in-domain} setting, whereas testing on the other datasets evaluates in the {\em out-of-domain} setting.  To test the union approach in the out-of-domain setting, we also evaluate the union with the training examples from the target test dataset excluded (S2AND (-target)).  We perform five runs with different random seeds
and report the average results in Table \ref{tab:transfer_b3} and Table \ref{tab:transfer_roc}.

The results for clustering are shown in Table \ref{tab:transfer_b3}, and the pairwise classification results are shown in Table \ref{tab:transfer_roc}.
As discussed earlier, each of the existing AND datasets tends to focus on only a subset of the scholarly literature, and how representative any of the datasets is of the real-world AND task faced by digital libraries is unclear.  Thus, in order to better gauge how systems are likely to perform on the real-world AND task faced in practice, our primary evaluation focuses on the out-of-domain setting, in which the test data distribution may differ from the training data distribution.  To measure out-of-domain performance, we evaluate in the setting where the training set for the target test set is not included in the training data (listed as ``S2AND (-target)'' in the table).
Compared to each out-of-domain single dataset (the not italicized values in Table \ref{tab:transfer_b3}), ``S2AND (-target)'' outperforms or matches all others on 4 out of 7 datasets, and is never further than 0.016 $B^3$ from the best. This indicates that training on the union of all datasets (excluding target) is an effective way to transfer to out-of-domain data, relative to training on a single one of the existing datasets. As mentioned in Section \ref{sec:s2and-data}, there is some overlap of records between datasets, and it is less than 1\% on average (relative to the size of each target dataset). Moreover, we also found an insignificant positive correlation between overlap proportions and transfer results in Table \ref{tab:transfer_b3}.  Both of these results suggest that the positive transfer in the out-of-domain setting is not due to simple overlap of the records in the data sets.

Regarding the in-domain setting, compared to each in-domain single dataset (the italicized values in Table \ref{tab:transfer_b3}), ``S2AND (+target)'' outperforms on 3 out 7 datasets, and is never more than 0.015 $B^3$ lower.  This indicates that training on the union of all datasets is an effective way to train a model that is robust across all the existing datasets.

For pairwise classification performance, we see similar trends, with the exception that Inspire is also best on 3 out of 8 datasets. 
S2AND (-target) is also best on 3 out of 8 datasets, and is never further than 0.026 AUROC from the best. In the in-domain setting, compared to each single dataset, S2AND (+target) outperforms or matches on 4 out 8 datasets, and is never further than 0.016 AUROC off.  We do not compute statistical significance for individual results here since the number of test examples is relatively large (the median size of a test set for one random seed is 1,436 examples, and we report the average of five seeds), and we expect our general findings to be reliable.

\subsubsection{Feature Importance}
We study the importance of individual features using SHapley Additive exPlanations (SHAP) \cite{lundberg2017unified} of the S2AND (-target) union model. SHAP values provide individualized, per-sample and per-feature linear additive explanations. Intuitively, a SHAP value $s_{ij}$ for sample $i$ and feature $j$ is a measure of the effect of this feature's presence (vs absence) on the model output for sample $j$, which in our case is the probability that two records are written by the same author. Upon manual examination of the SHAP plots for each of the seven S2AND (-target) models we observe that some features are nearly always the most important (last name counts and SPECTER similarities are the most notable examples, the latter of which is never lower than 4th most important), while other features have a more variable importance. For example (a) Jaro-Winkler, Levenshtein, and prefix first name similarities, (b) middle initials overlap, and (c) venue similarity are only near the top of the SHAP ordering for some of the datasets. 
\begin{table}
\begin{small}
\caption{\label{tab:main_ablation} Ablation experiments studying the effects of removing single features; substituting different clustering linkage functions (vs. the average linkage), clustering methods (vs. hierarchical agglomerative clustering), or pairwise classifiers (vs. gradient-boosted trees); or removing the monotonicity constraints or nameless classifier.  Most design alternatives hurt performance.}
\scalebox{0.97}{%
 \begin{tabular}{l l l} 
 \hline
 Ablation experiment & Average $B^3$ & $\Delta$ from baseline\\ [0.5ex] 
 \hline
Baseline (our final model) & 0.900 & -\\ \hline
Ward linkage               & 0.821 & 0.079 \\
Complete linkage           & 0.842 & 0.058 \\
Linear pairwise classifier & 0.870 & 0.030 \\
Single linkage             & 0.878 & 0.022 \\ 
DBSCAN clustering          & 0.878 & 0.022 \\
No specter                 & 0.879 & 0.021 \\
No affiliation             & 0.880 & 0.020 \\
No name counts             & 0.881 & 0.019 \\
No coauthor                & 0.883 & 0.017 \\
No nameless classifier     & 0.886 & 0.014 \\
No advanced names          & 0.893 & 0.007 \\
No references              & 0.895 & 0.005 \\
No email                   & 0.896 & 0.004 \\
No title                   & 0.897 & 0.003 \\
No venue, journal          & 0.897 & 0.003 \\
No year                    & 0.899 & 0.001 \\
No monotonicity            & 0.901 & -0.001 \\
\hline
\end{tabular}}
\end{small}
\vspace{-10pt}
\end{table}

\begin{table}
\begin{small}
\caption{\label{tab:second_ablation} Performance for different training set sizes.  S2AND's performance increases marginally with training set size.}
\scalebox{0.97}{%
 \begin{tabular}{l l l} 
 \hline
 Ablation experiment & Average $B^3$ & $\Delta$ from baseline\\ [0.5ex] 
 \hline
100k training (our final model) & 0.900 & - \\
 \hline
1k training  & 0.890 & 0.010 \\
5k training  & 0.895 & 0.005 \\
10k training & 0.900 & 0.000 \\
50k training & 0.899 & 0.001 \\
 \hline
\end{tabular}}
\end{small}
\vspace{-10pt}
\end{table}

\subsubsection{Ablation study} We now analyze the design choices and feature set of S2AND through ablation studies.  In these experiments we evaluate for five random seeds and report the mean (over random seeds) $B^3$ F1 performance of each variant averaged over the seven test sets in the S2AND (-target) setting. Table \ref{tab:main_ablation} shows the results of removing single features from S2AND, or by altering design choices in the classifier or clusterer.  The results show that many of the choices in S2AND are important for performance, in particular the use of average-linkage in the clusterer, the use of GBTs rather than a linear model, and the incorporation of SPECTER.  The three most valuable features are SPECTER embedding distance, affiliation, and name counts. Table \ref{tab:second_ablation} shows the effect of training size (of each dataset in the union) of the pairwise models. Performance increases only marginally between 5k examples and the full 100k examples used in S2AND. Although some of the features had small incremental $B^3$ improvements, we observed substantial qualitative improvements during our extensive hands-on evaluations when using both monotonicity constraints and the nameless classifier.

\subsection{Comparison with Semantic Scholar}
\label{sec:comps2}
As we mentioned in the introduction, one of the major goals of this work is to improve the name disambiguation system in a real-world bibliographic database, in this case Semantic Scholar. Here we compare the performance of our proposed system against the current \texttt{S2} clusters on the production website. Even though S2AND's feature values are computed from \texttt{S2}'s data, \texttt{S2}'s AND system is very different from our reference AND algorithm trained on S2AND.  \texttt{S2}'s algorithm relies on hand-crafted rules and was tuned using a single dataset based on ORCID.  By contrast, our system is a learned model trained on the diverse datasets in S2AND. 

In addition to the overall performance, we also compare performance broken out across facets to measure whether overall accuracy and S2AND's improvements differ across different author or paper characteristics.  For fair comparison with Semantic Scholar, which was not trained on any of the datasets we consider here, we evaluate S2AND in the ``union excluding target'' setting.

\begin{figure*}[t!]
    \centering\includegraphics[scale=0.35]{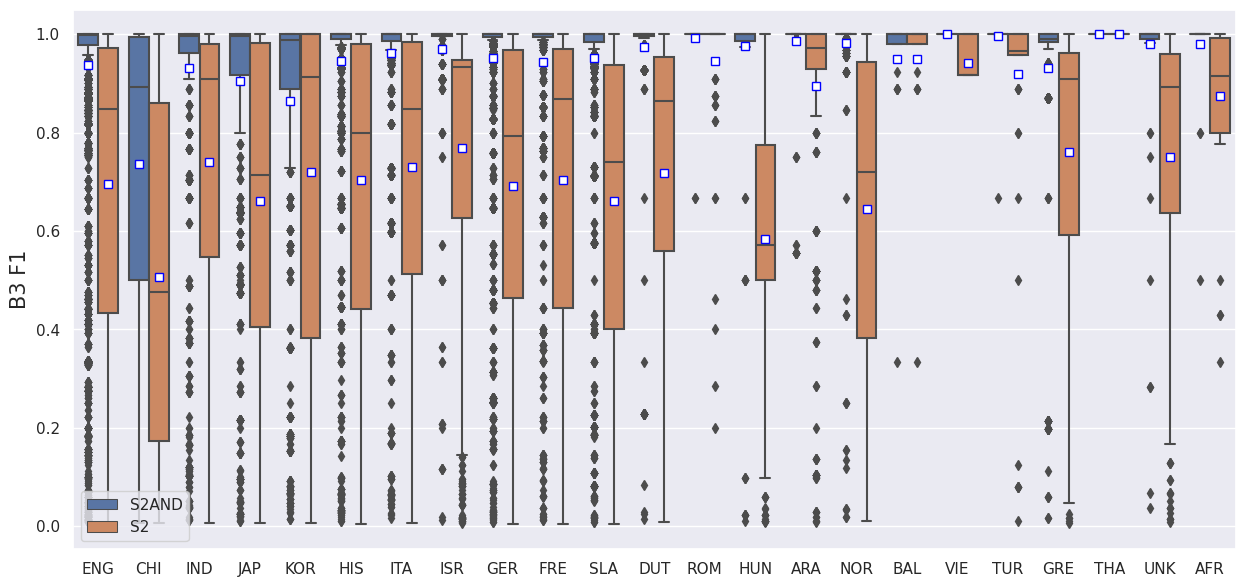}
    \vspace{-10pt} 
    \caption{Performance across estimated name geographic origin groups.  The performance of both systems varies significantly across different groups, and S2AND improves performance significantly over \texttt{S2} for almost all groups, and the improvement is relatively larger for groups with lower \texttt{S2} performance.}
    \label{fig:ethnicity}
    \vspace{-5pt}
 \end{figure*}

Overall, S2AND reduces error over \texttt{S2} by more than 50\%, as shown in Table \ref{tab:transfer_b3}, where S2AND's average $B^3$ F1 is 90\% compared to \texttt{S2} at 78.4\%.  The $B^3$ F1 performance broken out by estimated name geographic origin is shown in Figure \ref{fig:ethnicity}.  As discussed above, the proxy we use as facet values for this analysis is a highly imperfect indicator of the actual attributes of the author.  Nonetheless, the fact that AND performance of both systems varies significantly across the groups suggests areas for improvement in AND systems, and it is perhaps encouraging that S2AND tends to improve performance relatively more for groups for which S2's original performance was lower.  

\begin{figure*}[t!]
    \centering
    \begin{subfigure}
        \centering
        \includegraphics[height=1.825in]{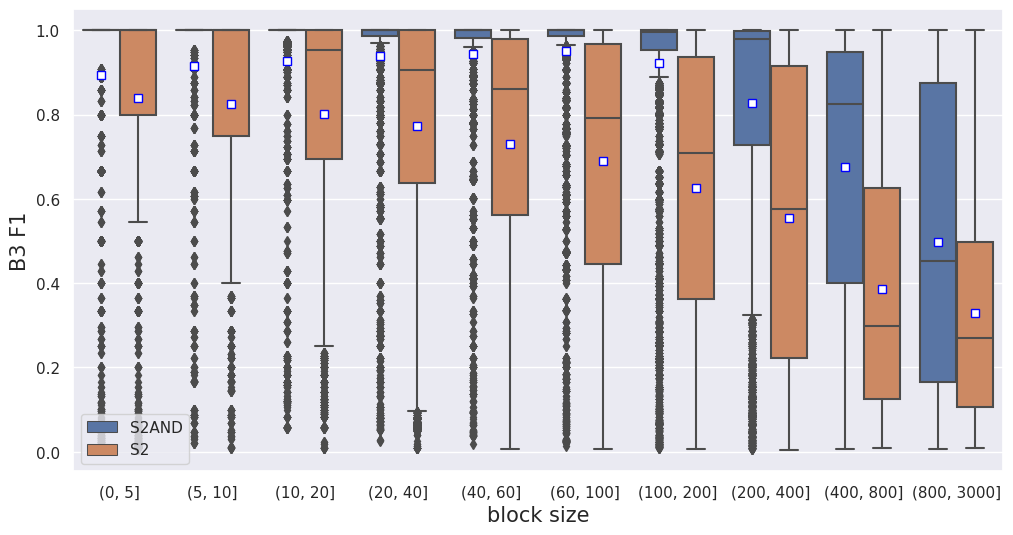}
    \end{subfigure}%
    ~ 
    \begin{subfigure}
        \centering
        \includegraphics[height=1.825in]{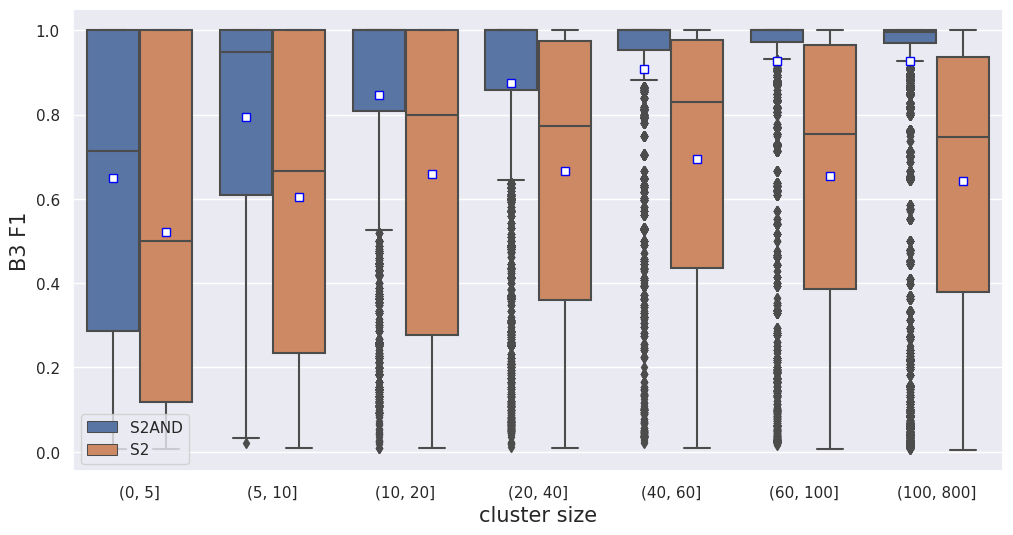}
    \end{subfigure}
    \vspace{-10pt} 
    \caption{Performance across (a) block sizes and (b) cluster sizes (number of papers by author). Larger blocks are more difficult for both algorithms, and S2AND improves performance across all block sizes. S2AND improves performance across all cluster sizes, with a larger gain for more prolific authors.}
    \label{fig:block_and_cluster_sizes}
    \vspace{-5pt}
\end{figure*}

\begin{figure*}[t!]
    \centering
    \begin{subfigure}
        \centering
        \includegraphics[height=1.825in]{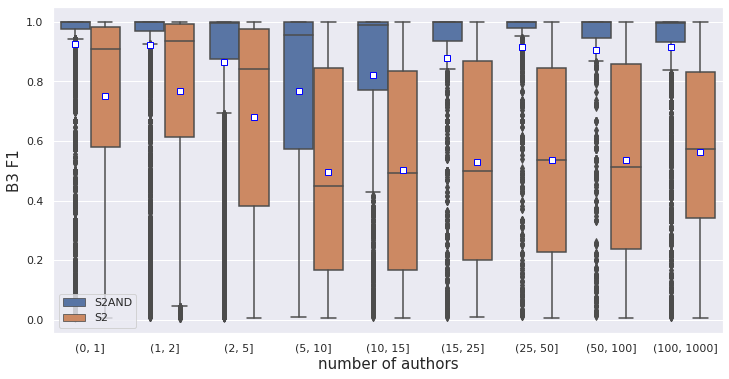}
    \end{subfigure}%
    ~ 
    \begin{subfigure}
        \centering
        \includegraphics[height=1.825in]{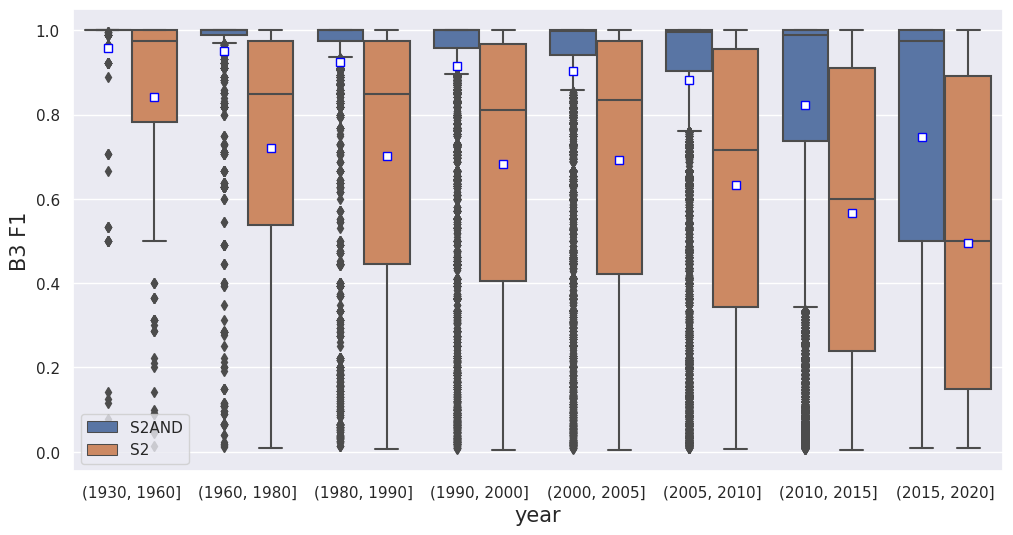}
    \end{subfigure}
    \vspace{-20pt} 
    \caption{Performance across (a) number of authors in a paper and (b) publication years.  S2AND improves performance over \texttt{S2} for any number of authors, with 5- to 10-author papers showing the lowest performance for both algorithms. S2AND improves performance for all years, and newer papers are more challenging for both algorithms in our data.}
    \label{fig:num_authors_and_pub_years}
    \vspace{-5pt}
\end{figure*}

\begin{figure*}[t!]
    \centering
    \begin{subfigure}
        \centering
        \includegraphics[height=1.81in]{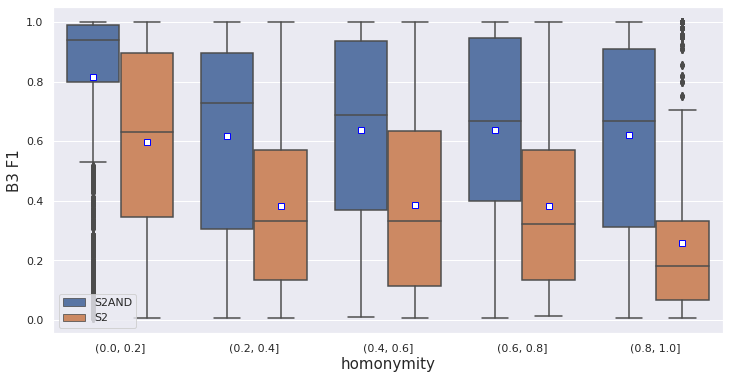}
    \end{subfigure}%
    ~ 
    \begin{subfigure}
        \centering
        \includegraphics[height=1.81in]{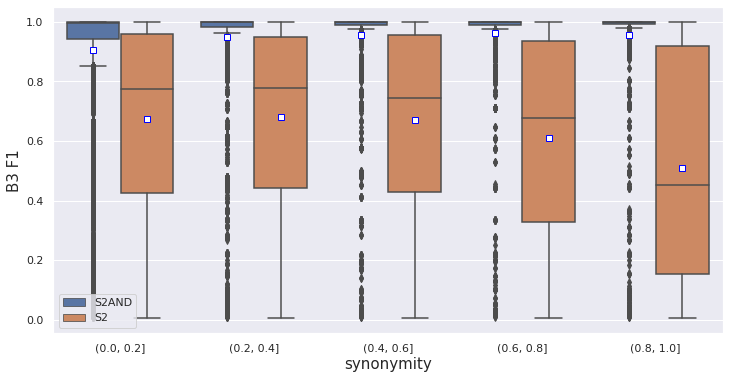}
    \end{subfigure}
    \vspace{-20pt} 
    \caption{Performance across varying author name (a) homonymity (higher values indicate more ambiguous names within a particular block) and (b) synonymity (higher values indicate more distinct name variants for a single author within a particular block).  S2AND improves performance in all cases.}
    \label{fig:homonym_and_synonym}
    \vspace{-5pt}
\end{figure*}

We present additional analyses in Figures \ref{fig:block_and_cluster_sizes}-\ref{fig:homonym_and_synonym}, isolating performance for the number of papers published by the author, along with attributes of the paper including the year of publication and the number of authors.  Finally, we also consider attributes of the block, including its size and two ratios that characterize its difficulty: {\em homonymity}, defined as the fraction of records in the block with the same names that are in different clusters, and its {\em synonymity}, the fraction of records in the same cluster that have different names.  In general, the results show that S2AND consistently improves performance across the different facet values.  Also, S2AND tends to show a larger improvement in cases where the original \texttt{S2} performance was lower.  Among these findings, one is that authors on newer papers can be more difficult for the systems to disambiguate---we attribute this in large part to the fact that the most recent papers in the challenging Aminer dataset are more often by junior authors with smaller clusters, and thus have higher error rates.  Further, one important area for improvement, highlighted in Figure \ref{fig:block_and_cluster_sizes}, involves disambiguating larger blocks.

\vspace{-5pt}
\section{Practical considerations for AND}
We also investigated building a production AND system out of the S2AND-trained reference model evaluated in the previous section.  As is common when applying machine learning to a real-world task, optimizing a metric on supervised datasets does not fully capture all the performance requirements of an actual AND system. Here we briefly comment on a handful of tactics we used for building out our production system. First, real data contains many ``paper" records that are not actually papers, and many papers that are not in English. Our features, particularly SPECTER embeddings, are not intended for use on such records. Second, the distribution of missing metadata in a real corpus does not perfectly match that of the datasets in this paper, and if one piece of metadata being missing is strongly correlated with a particular training dataset, the model can learn a spurious interaction and make counter-intuitive predictions. Third, not all errors are created equal. Errors that are obvious to a human are both easier to identify, and more likely to incur user complaints. To partially address these issues, we:
\begin{itemize}
\item Trained with an additional dataset, created by taking the union of all the other datasets and randomly knocking out features (e.g. randomly remove affiliation information). This helps harden the model against inappropriate predictions with missing metadata.
\item Added simple rules that prevent incompatible names from clustering together (e.g. John cannot cluster with James). These rules are not perfect, as authors can change their names, and names can be mis-extracted from papers, but they help us prevent the model from making many obvious errors which break user trust.
\item Assessed not only with $B^3$ on held-out data but also via qualitative evaluations on author profile corrections made by users to our previous AND model. We also extensively studied the outputs of many model iterations and adjusted the feature computations based on our observations.
\end{itemize}

\noindent There are also practical difficulties that we did not fully address: (a) recovering from name blocking errors, (b) lossy transliteration of, for example, Chinese names, (c) successfully clustering English and non-English papers together, and (d) the hierarchical agglomerative clustering pipeline does not allow for the similarity of one pair of records to influence the similarity of another pair of records.

\section{Conclusion}

We have presented S2AND, a new dataset and benchmark for author disambiguation that unifies eight previous AND datasets into a uniform format.  In experiments with a reference AND implementation that we introduce, we show that training on the union of datasets in S2AND improves generalization to datasets not seen in training.  Our benchmark algorithm also improves over the production Semantic Scholar system, reducing error by more than 50\%. We hope that S2AND helps further new innovation in the important task of author disambiguation in digital libraries.

\section{Acknowledgments}
We thank Zejiang Shen, Sonia Murthy, Kyle Lo, and Dan Weld for helpful feedback; Bailey Kuehl and Rodney Kinney for their extensive evaluation; Regan Huff, Jason Dunkelberger, Joanna Power, Angele Zamarron, and Brandon Stilson for their engineering work that turned our Python code into something that scales to 200 million papers; and the entire Semantic Scholar team for creating the underlying data.  This work was supported in part by NSF grant OIA-2033558.

\bibliographystyle{IEEEtranN}
\bibliography{references}

\end{document}